\documentclass[aps,prd,twocolumn,
comment, 
superscriptaddress,groupedaddress,
amsmath,amssymb,floatfix,nofootinbib]
{revtex4}  
\usepackage{float}
\usepackage{graphicx}
\usepackage{bm}
\usepackage[usenames,dvipsnames]{color}
\usepackage{slashed}
\usepackage{slashed}

\begin{document}

%

\let\a=\alpha      \let\b=\beta       \let\c=\chi        \let\d=\delta
\let\e=\varepsilon \let\f=\varphi     \let\g=\gamma      \let\h=\eta
\let\k=\kappa      \let\l=\lambda     \let\m=\mu
\let\o=\omega      \let\r=\varrho     \let\s=\sigma
\let\t=\tau        \let\th=\vartheta  \let\y=\upsilon    \let\x=\xi
\let\z=\zeta       \let\io=\iota      \let\vp=\varpi     \let\ro=\rho
\let\ph=\phi       \let\ep=\epsilon   \let\te=\theta
\let\n=\nu
\let\D=\Delta   \let\F=\Phi    \let\G=\Gamma  \let\L=\Lambda
\let\O=\Omega   \let\P=\Pi     \let\Ps=\Psi   \let\Si=\Sigma
\let\Th=\Theta  \let\X=\Xi     \let\Y=\Upsilon
%


\def\cA{{\cal A}}                \def\cB{{\cal B}}
\def\cC{{\cal C}}                \def\cD{{\cal D}}
\def\cE{{\cal E}}                \def\cF{{\cal F}}
\def\cG{{\cal G}}                \def\cH{{\cal H}}
\def\cI{{\cal I}}                \def\cJ{{\cal J}}
\def\cK{{\cal K}}                \def\cL{{\cal L}}
\def\cM{{\cal M}}                \def\cN{{\cal N}}
\def\cO{{\cal O}}                \def\cP{{\cal P}}
\def\cQ{{\cal Q}}                \def\cR{{\cal R}}
\def\cS{{\cal S}}                \def\cT{{\cal T}}
\def\cU{{\cal U}}                \def\cV{{\cal V}}
\def\cW{{\cal W}}                \def\cX{{\cal X}}
\def\cY{{\cal Y}}                \def\cZ{{\cal Z}}


\def\be{\begin{equation}}
\def\ee{\end{equation}}
\def\bea{\begin{eqnarray}}
\def\eea{\end{eqnarray}}
\def\bm{\begin{matrix}}
\def\em{\end{matrix}}
\def\bpm{\begin{pmatrix}}
    \def\epm{\end{pmatrix}}

{\newcommand{\lsim}{\mbox{\raisebox{-.6ex}{~$\stackrel{<}{\sim}$~}}}
{\newcommand{\gsim}{\mbox{\raisebox{-.6ex}{~$\stackrel{>}{\sim}$~}}}
\def\mpl{M_{\rm {Pl}}}
\def\gev{{\rm \,Ge\kern-0.125em V}}
\def\tev{{\rm \,Te\kern-0.125em V}}
\def\mev{{\rm \,Me\kern-0.125em V}}
\def\ev{\,{\rm eV}}

\title{\boldmath  $\bar{B}\rightarrow D^{(\ast)}\tau \bar{\nu}$ excesses in ALRSM constrained from $B$, $D$ decays and $D^{0}-\bar{D}^{0}$ mixing}
\author{Chandan Hati}
\email{chandan@prl.res.in} 
\affiliation{Physical Research Laboratory, Navrangpura, Ahmedabad 380 009, India}
\affiliation{Indian Institute of Technology Gandhinagar, Chandkheda, Ahmedabad 382 424, India}
\author{Girish Kumar}
\email{girishk@prl.res.in}
\affiliation{Physical Research Laboratory, Navrangpura, Ahmedabad 380 009, India}
\affiliation{Indian Institute of Technology Gandhinagar, Chandkheda, Ahmedabad 382 424, India}
\author{Namit Mahajan}
\email{nmahajan@prl.res.in} 
\affiliation{Physical Research Laboratory, Navrangpura, Ahmedabad 380 009, India}

\begin{abstract}
Recent experimental results from the LHCb, BaBar and Belle collaborations on the semitauonic decays of $B$ meson, $\bar{B}\rightarrow D^{(\ast)}\tau \bar{\nu}$, showing a significant deviation from the Standard Model (SM), hint towards a new physics scenario beyond the SM. In this work, we show that these enhanced decay rates can be explained within the framework of $E_{6}$ motivated Alternative Left-Right Symmetric Model (ALRSM), which has been successful in explaining the recent CMS excesses and has the feature of accommodating high scale leptogenesis. The R-parity conserving couplings in ALRSM can contribute universally to both $\bar{B}\rightarrow D \tau \bar{\nu}$ and $\bar{B}\rightarrow D^{(\ast)}\tau \bar{\nu}$ via the exchange of scalar leptoquarks. We study the leptonic decays $D_{s}^{+} \rightarrow \tau^{+} \bar{\nu}$, $B^{+}\rightarrow \tau^{+} \bar{\nu}$, $D^{+}\rightarrow \tau^{+} \bar{\nu}$ and $D^{0}$-$\bar{D}^{0}$ mixing to constrain the couplings involved in explaining the enhanced $B$ decay rates and we find that ALRSM can explain the current experimental data on $\cR(D^{(\ast)})$ quite well while satisfying these constraints.

\end{abstract}
\maketitle

\section{Introduction}
Recently the LHCb collaboration has reported the ratio of branching fractions for the semitauonic decay of $B$ meson, $\bar{B}\rightarrow D^{\ast}\tau \bar{\nu}$, to be $\cR(D^{\ast})=0.336\pm 0.027 ({\rm stat.}) \pm 0.030 ({\rm syst.})$ with the Standard Model (SM) expectation $0.252\pm0.005$, amounting to a $2.1\sigma$ excess \cite{Aaij:2015yra}. In general, the observables are introduced as ratios to reduce theoretical uncertainties 
\be{\label{1.1}}
\cR(X)=\frac{\cB(\bar{B}\rightarrow X\tau \bar{\nu})}{ \cB(\bar{B}\rightarrow X l \bar{\nu})},
\ee
where $l=e, \mu$. This measurement is in agreement with the measurements of $\bar{B}\rightarrow D^{(\ast)}\tau \bar{\nu}$ reported by the BaBar \cite{Lees:2012xj, Lees:2013uzd} and Belle \cite{Huschle:2015rga} collaborations. The results reported by BaBar and Belle are given by $\cR(D)^{{\rm BaBar}}=0.440\pm 0.058  \pm 0.042 $, $\cR(D)^{{\rm Belle}}=0.375\pm 0.064  \pm 0.026 $ and $\cR(D^{\ast})^{{\rm BaBar}}=0.332\pm 0.024 \pm 0.018 $, $\cR(D^{\ast})^{{\rm Belle}}=0.293\pm 0.038  \pm 0.015 $, with the SM expectations given by $\cR(D)^{{\rm SM}}=0.300\pm 0.010$ and $\cR(D^{\ast})^{{\rm SM}}=0.252\pm 0.005$. These results are consistent with earlier measurements \cite{Aubert:2007dsa, Bozek:2010xy} and when combined together show a significant deviation from the SM. 

Several new physics (NP) scenarios accommodating semileptonic $b\rightarrow c$ decay  have been proposed to explain these excesses. The two-Higgs Doublet Model (2HDM) of type II is one of the well studied candidates of NP which can affect the semitauonic $B$ decays significantly \cite{Krawczyk:1987zj, Kalinowski:1990ba, Hou:1992sy, Tanaka:1994ay, Kamenik:2008tj, Nierste:2008qe, Tanaka:2010se}. However, the BABAR collaboration has excluded the 2HDM of type II at 99.8 \% confidence level \cite{Lees:2012xj, Lees:2013uzd}. Phenomenological studies of the four fermion operators that can explain the discrepancy have been addressed in Refs. \cite{Fajfer:2012vx, Sakaki:2012ft, Fajfer:2012jt, Datta:2012qk, Bailey:2012jg, Becirevic:2012jf, Tanaka:2012nw, Freytsis:2015qca, Bhattacharya:2015ida}. The excesses have been explained in a more generalized framework of 2HDM in Refs. \cite{Celis:2012dk, Ko:2012sv, Crivellin:2012ye} and in the framework of $R$-parity violating (RPV) Minimal Supersymmetric Standard Model (MSSM) in Ref. \cite{Deshpande:2012rr}. While in Refs. \cite{Dorsner:2009cu,Fajfer:2012jt, Tanaka:2012nw, Freytsis:2015qca, Sakaki:2013bfa} the excesses have been addressed in the context of leptoquark models. In Ref. \cite{Greljo:2015mma}, a dynamical model based on a $SU(2)_{L}$ triplet of massive vector bosons, with predominant coupling to third generation fermion was proposed to explain the excesses, while other alternative approaches have been taken in Refs.  \cite{Dorsner:2013tla, Biancofiore:2013ki, Fan:2015kna}. 

From a theoretical point of view, NP scenarios explaining the above discrepancies and addressing other direct or indirect collider searches for NP are particularly intriguing. To this end, we must mention the recently announced results for the right-handed gauge boson $W_{R}$ search at $\sqrt{s}=8 \rm{TeV}$ and $19.7 \rm{fb}^{-1}$ of integrated luminosity by the CMS Collaboration at the LHC. They have reported 14 observed events with 4 expected SM background events, amounting to a $2.8\sigma$ local excess in the bin $1.8 \tev< m_{eejj}<2.2 \tev$, which cannot be explained in the standard framework of Left-Right Symmetric Model (LRSM) with the gauge couplings $g_{L}=g_{R}$ \cite{Khachatryan:2014dka}.  On the other hand, the CMS search for di-leptoquark production at $\sqrt{s}=8 \rm{TeV}$ and $19.6 \rm{fb}^{-1}$ of integrated luminosity have been reported to show a $2.4\sigma$ in the $eejj$ channel and a $2.6\sigma$ local excess in the $e\slashed{p}_{T}jj$ channel corresponding to $36$ observed events with $20.49\pm 2.4\pm 2.45$(syst.) expected SM events in the $eejj$ channel and $18$ observed events with $7.54\pm 1.20\pm 1.07$(syst.) expected SM events in the $e\slashed{p}_{T}jj$ channel respectively \cite{CMS:2014qpa}. These excesses has been explained from $W_R$ decay in the framework of LRSM with $g_L\ne g_R$ embedded in the $SO(10)$ gauge group in  Refs. \cite{Deppisch:2014qpa, Deppisch:2014zta, Dev:2015pga} and in LRSM with $g_L = g_R$ by taking into account the CP phases and non-degenerate masses of heavy neutrinos in Ref. \cite{Gluza:2015goa}, while other NP scenarios have been proposed in Refs. \cite{Dobrescu:2014esa, Aguilar-Saavedra:2014ola, Allanach:2014lca, Queiroz:2014pra, Biswas:2014gga, Allanach:2014nna, Allanach:2015ria, Dhuria:2015hta, Dutta:2015osa, Dobrescu:2015asa, Berger:2015qra, Krauss:2015nba, Dhuria:2015swa}. Interestingly, in some of these NP scenarios attempts were made to explain the discrepancies in decays of $B$ meson in an unified framework \cite{Biswas:2014gga} or separately \cite{Deshpande:2012rr}.

In this paper we study the flavor structure of the $E_{6}$ motivated Alternative Left-Right Symmetric Model (ALRSM) \cite{Ma:1986we}, which can explain the CMS excesses and accommodate high scale leptogenesis \footnote{Note that in the conventional LRSM framework the canonical mechanism of leptogenesis is inconsistent with the range of $W_R$ mass $(\sim 2 \tev)$ corresponding to the excess at CMS \cite{Dhuria:2015wwa, Dhuria:2015cfa}.} \cite{Dhuria:2015hta}, to explore if this framework can address the experimental data for $\cR(D^{(\ast)})$ explaining the discrepancy with the SM expectations. This scenario is particularly interesting because unlike the R-parity violating MSSM in Refs. \cite{Allanach:2014lca, Biswas:2014gga, Deshpande:2012rr}, this model involves only R-parity conserving interactions. Furthermore, a careful analysis of the flavor physics constraints, such as the rare decays and the mixing of mesons can play a crucial role in determining the viability of any NP scenario. Therefore, we study the leptonic decays $D_{s}^{+} \rightarrow \tau^{+} \bar{\nu}$, $B^{+}\rightarrow \tau^{+} \bar{\nu}$, $D^{+}\rightarrow \tau^{+} \bar{\nu}$ and $D^{0}$-$\bar{D}^{0}$ mixing to constrain the semileptonic $b \rightarrow c$ transition in ALRSM. We find that despite being constrained by the above processes ALRSM can explain the current experimental data on $\cR(D^{(\ast)})$ quite well.

The rest of this article is organized as follows. In Sec. \ref{sec1}, we discuss the effective Hamiltonian and the general four-fermion operators that can explain the $\cR(D^{(\ast)})$ data. In Sec. \ref{sec2}, we introduce ALRSM and present the viable interactions, followed by the evaluation of the Wilson coefficients. In Sec. \ref{sec3}, we discuss the constrains from the leptonic decays $D_{s}^{+}\rightarrow \tau^{+} \bar{\nu}$, $B^{+}\rightarrow \tau^{+} \bar{\nu}$, $D^{+}\rightarrow \tau^{+} \bar{\nu}$ and mixing between $D^{0}$-$\bar{D}^{0}$. In Sec. \ref{sec4}, we summarize our results and conclude.
\section{The effective Hamilonian for $\bar{B}\rightarrow D^{(\ast)}\tau \bar{\nu}$ decay}
\label{sec1}
To include the effects of NP, the SM effective Hamiltonian for the quark level transition $b\rightarrow c l \bar{\nu}_l$ can be augmented with a set of four-Fermi operators in the following form \cite{Sakaki:2012ft}
\begin{widetext}
\be
\label{3.0}
 \mathcal{H}_{eff} = \frac{4 G_F}{\sqrt{2}} V_{cb} \sum_{l= e,\mu,\tau}[(1+C_{V_L}^l)O_{V_L}^l + C_{V_R}^l O_{V_R}^l + C_{S_L}^l O_{S_L}^l + C_{S_R}^l O_{S_R}^l + C_{T_L}^l O_{T_L}^l],
\ee
\end{widetext}
where $G_F$ is the Fermi constant, $V_{cb}$ is the appropriate CKM matrix element and $C^l_i$ (i = $V_{L/R}$, $S_{L/R}$, $T_L$) are the Wilson coefficients associated with the new effective vector, scalar and tensor interaction operators respectively. These new six dimensional four-Fermi operators are generated by NP at some energy higher than the electroweak scale and are defined as
\bea \label{1.1.1}
O_{V_L}^l &=& (\bar{c}_L\gamma^\mu b_L) (\bar{l}_L\gamma_\mu\nu_{l L}),\nonumber\\
O_{V_R}^l &=& (\bar{c}_R\gamma^\mu b_R) (\bar{l}_L\gamma_\mu\nu_{l L}),\nonumber\\
O_{S_L}^l &=& (\bar{c}_R b_L) (\bar{l}_R\nu_{l L}),\nonumber\\
O_{S_R}^l &=& (\bar{c}_L b_R) (\bar{l}_R\nu_{l L}),\nonumber\\
O_{T_L}^l &=& (\bar{c}_R\sigma^{\mu\nu} b_L) (\bar{l}_R\sigma_{\mu\nu}\nu_{l L}),
\eea
where $\sigma^{\mu\nu} = (i/2)[\gamma^\mu,\gamma^\nu]$. The SM effective Hamiltonian corresponds to the case $C^l_i = 0$. Note that in writing the general $ \mathcal{H}_{eff}$, we have neglected the tiny contributions from the right-handed neutrinos and therefore, we treat the neutrinos to be left-handed only.\\
In order to perform the numerical analysis of the transition $B\rightarrow D^{(*)}\tau \nu$, we need to have the knowledge of the
hadronic form factors which parametrize the vector, scalar and tensor current matrix elements.  The $B \rightarrow D^{(\ast)}\tau \nu$ matrix elements of the aforementioned effective operators depend on the momentum transfer between B and $D^{(\ast)} (q^\mu = p_B^\mu -k^\mu)$ and   are generally parametrized in the following way \cite{Sakaki:2012ft, Hagiwara:1989cu}
\begin{widetext}
\bea
\label{3.1}
\langle D(k)\lvert \bar{c}\gamma_\mu b\rvert \bar{B}(p_B)\rangle &=& \left[(p_B + k)_\mu -\frac{m_B^2 - m_D^2}{q^2} q_\mu \right] F_1(q^2) + q_\mu \frac{m_B^2 - m_D^2}{q^2} F_0(q^2),\\
\langle D^{\ast}(k,\epsilon)\lvert \bar{c}\gamma_\mu b\rvert \bar{B}(p_B)\rangle &=& -i \epsilon_{\mu\nu\rho\sigma}\epsilon^{\nu \ast}p_B^{\rho}k^{\sigma}\frac{2 V(q^2)}{m_B+m_{D^{\ast}}},\\
\langle D^{\ast}(k,\epsilon)\lvert \bar{c}\gamma_\mu\gamma_5 b\rvert \bar{B}(p_B)\rangle &=& \epsilon^{\mu \ast} (m_B + m_{D^{\ast}})A_1(q^2) - (p_B+k)_\mu(\epsilon^{\ast}\cdot q)\frac{A_2(q^2)}{m_B +m_{D^{ast}}} \nonumber\\ 
&& -q_\mu(\epsilon^{\ast}\cdot q) \frac{2 m_{D^{\ast}}}{q^2}\left(A_3(q^2)-A_0(q^2)\right),
\eea
\bea
\langle D^{*}(k,\epsilon)\lvert \bar{c}\sigma_{\mu\nu} b\rvert \bar{B}(p_B)\rangle &=&
\epsilon_{\mu\nu\rho\sigma}\left\{-\epsilon^{\ast \rho}(p_B+k)^\sigma T_1(q^2)+\epsilon^{\ast \rho} q^{\sigma}\frac{m_B^2-m_{D^{\ast}}}{q^2}(T_1(q^2)-T_2(q^2))\right. \nonumber\\
&&+ \left. 2 \frac{{\epsilon^{\ast} q}}{q^2} p_B^\rho k^\sigma\left(T_1(q^2) - T_2(q^2) -\frac{q^2}{m_{B^2}-m_{D^{\ast 2}}}T_3(q^2)\right)\right\},
\eea
\end{widetext}
where $F_1(0) = F_0(0)$, $A_3(0) = A_0(0)$ and 
\be
\label{3.2}
A_3(q^2) = \frac{m_B + m_{D^{\ast}}}{2 m_{D^\ast}}A_1(q^2) - \frac{m_B - m_{D^{\ast}}}{2 m_{D^\ast}}A_2(q^2).
\ee 
$\epsilon_\mu$ is the polarization vector of the $D^\ast$. Note that the hadronic matrix elements of the scalar and pseudoscalar operators can be conveniently derived from their vector counterpart by applying the equations of motion $-i\partial^\mu(\bar{q}_a \gamma_\mu q_b) = (m_a-m_b)\bar{q}_aq_b$ and $-i\partial^\mu(\bar{q}_a\gamma_\mu\gamma_5 q_b) = (m_a+m_b)\bar{q}_a \gamma_5 q_b$. However, in what follows, we choose to work with the following parametrization of the form factors which are more suitable for including the results of the heavy quark effective theory (HQET). The matrix elements of the vector and axial vector operators can be expressed as \cite{Tanaka:1994ay, Neubert:1991td}
\bea
\label{3.3}
\langle D(v')\lvert\bar{c}\gamma_\mu b\rvert\bar{B}(v)\rangle &=& \sqrt{m_B m_D}\left\{\xi_{+}(w)(v+v')_\mu \right. \nonumber \\
&& \left. + \xi_{-}(w)(v-v')_\mu \right\}\nonumber\\
\nonumber\\
\langle D^\ast(v',\epsilon)\lvert\bar{c}\gamma_\mu b\rvert\bar{B}(v)\rangle   &=&  i\sqrt{m_B m_{D^\ast}} \xi_V(w) \epsilon_{\mu\nu\rho\sigma}\epsilon^{\ast\nu}v'^\rho v^\sigma, \nonumber\\
 \nonumber\\
\langle D^\ast(v',\epsilon)\lvert\bar{c}\gamma_\mu \gamma_5 b\rvert\bar{B}(v)\rangle   &=& \sqrt{m_B m_{D^\ast}} \left\{\xi_{A_1}(w)(w+1)\epsilon_\mu^\ast \right. \nonumber \\
&&- (\epsilon^\ast\cdot v)\nonumber\\
&&\left.\left(\xi_{A_2}(w)v_\mu + \xi_{A_3}(w)v'^\mu\right) \right\}.
\eea
The form factors of tensor operators are defined as \cite{Tanaka:2012nw}
\bea
\label{3.4}
\langle D(v')\lvert\bar{c}\sigma_{\mu\nu} b\rvert\bar{B}(v)\rangle &=& -i \sqrt{m_B m_{D}} \xi_T(w) \left(v_\mu v'_\nu - v_\nu v'_\mu \right),\nonumber\\
\langle D^\ast(v')\lvert\bar{c}\sigma_{\mu\nu} b\rvert\bar{B}(v)\rangle &=& -i \sqrt{m_B m_{D^\ast}} \epsilon_{\mu\nu\rho\sigma}\nonumber\\
&&\left\{\xi_{T_1}(w)\epsilon^{\ast \rho}(v+v')^\rho \right.\nonumber\\
&&  + \left.\xi_{T_2}(w)\epsilon^{\ast \rho}(v-v')^\sigma  \right.\nonumber\\
&&  + \left.\xi_{T_3}(w)(\epsilon^\ast \cdot v)(v+v')^\rho (v-v')^\sigma \right\},\nonumber\\
\eea
where $v = p_B/m_B$ and $v' = k/m_{D^{(\ast)}}$ are the four-velocities of the B and $D^{(\ast)}$ mesons respectively, and the kinematic variable $w(q^2)$ is the product of the velocities of initial and final mesons $w(q^2) = \left(m_B^2 + m_{D^{(\ast)}} - q^2 \right)/2m_B m_{D^{\ast}}$. The HQET and QCD dispersive techniques can be used to constrain the shapes of these form factors \cite{Caprini:1997mu}. To this end, the HQET form factors are redefined as linear combinations of the different form factors $V_1(w)$, $S_1(w)$, $A_1(w)$ and $R_{1,2,3}(w)$ \cite{Caprini:1997mu, Tanaka:2012nw}, which reduces to the universal Isgur-Wise function \cite{Isgur:1989vq} normalized to unity at $w=1$ in the heavy quark limit. The form factors in the parameterization of  Caprini {\it et al.} \cite{Caprini:1997mu}, which describes the shape and normalization in terms of the four quantities: the normalizations $V_1(1)$, $A_1(1)$, the slopes $\rho^2_D$, $\rho^2_{D^\ast}$ and the amplitude ratios $R_1(1)$ and $R_2(1)$ (determined by measuring the differential decay width as a function of $w$). The form factors $V_1(w)$ and $S_1(w)$ contribute to the decay $B\rightarrow D l \bar{\nu}_l$ $(l=e, \mu, \tau)$, while the decay $B\rightarrow D^\ast l \bar{\nu}_l$ receives contributions from $A_1(w)$ and $R_{1,2,3}(w)$. However, the semileptonic decay into light charged leptons $B\rightarrow D l \bar{\nu}_l$ involves only $V_1(w)$ and therefore, $V_1(w)$ can be measured experimentally. The parametrization of the form factors in terms of the slope parameters $\rho^2_D$, $\rho^2_{D^\ast}$ and the value of the respective form factors at the kinematic end point $w=1$ is given by \cite{Caprini:1997mu, Neubert:1993mb} 
\bea
\label{3.5}
V_1(w) &=& V_1(1) \left\{1-8 \rho^2_D z + (51 \rho_D^2 - 10)z^2 \right.\nonumber\\
 && \left. - (252 \rho_D^2 - 84) z^3\right\},\\
 A_1(w) &=& A_1(1)\left\{1-8\rho^2_{D^\ast}z + (53\rho^2_{D^\ast}-15)z^2 \right.\nonumber\\ 
  && \left. -(231\rho^2_{D^\ast}-91)z^3 \right\},\\
  R_1(w) &=& R_1(1) - 0.12(w-1) + 0.05(w-1)^2,\nonumber\\
  R_2(w) &=& R_2(1) + 0.11(w-1) - 0.06(w-1)^2,\nonumber\\
  R_3(w) &=& 1.22 - 0.052(w-1) + 0.026(w-1)^2,\nonumber\\
 \eea
 with $z = (\sqrt{w+1}-\sqrt{2})/(\sqrt{w+1}+\sqrt{2})$. For $S_1(w)$ we use the parametrization given in Ref. \cite{Tanaka:2010se}
 
 \bea
 S_1(w) &=& V_1(w)\left\{1 + \Delta\left(-0.019 + 0.041 (w-1)\right.\right.\nonumber\\
 && \left.\left. -0.015(w-1)^2\right)\right\},
 \eea 
 
  with $\Delta = 1 \pm 1$. By fitting the measured quantity $|V_{cb}| V_1(w)$ to the two parameter ansatz as given in Eq.\eqref{3.5}, the heavy flavor averaging group (HFAG) extracts the following parameters: $V_1(1)|V_{cb}| = (42.65 \pm 1.53)\times 10^{-3}$, $\rho^2_D = 1.185 \pm 0.054$ \cite{Amhis:2014hma}. In the case of $B\rightarrow D^\ast l \bar{\nu}_l$, HFAG determines $A_1(1)|V_{cb}|=(35.81 \pm 0.45)\times 10^{-3}$, $\rho_{D^\ast}=1.207 \pm 0.026$, $R_1(1)=1.406 \pm 0.033$ and $R_2(1) = 0.853 \pm 0.020$ by performing a four-dimensional fit of the parameters \cite{Amhis:2014hma}. However, since the fitted curve are plagued with large statistical and systematic uncertainties, for form factor normalizations,  we use $V_1(1) = 1.081 \pm 0.024$ from the recent lattice QCD calculations \cite{Okamoto:2004xg} and  for $A_1(1)$ we use the updated value $A_1(1) = 0.920 \pm 0.014$ from the FNAL/MILC group \cite{Bailey:2014tva}. The amplitude ratios $R_1(1)$ and $R_2(1)$ are determined from the fit by HFAG $R_1(1)=1.406 \pm 0.033$, $R_2(1) = 0.853 \pm 0.020$ \cite{Amhis:2014hma}. 

\section{Alternative Left-Right Symmetric Model and analysis of operators mediating $\bar{B}\rightarrow D^{(\ast)}\tau \bar{\nu}$}
\label{sec2}
One of the maximal subgroups of superstring inspired $E_{6}$ group is given by $SU(3)_{C}\times SU(3)_{L} \times SU(3)_{R}$. The fundamental $27$ representation of $E_{6}$ can be decomposed under this subgroup as
\be{\label{4.0.0}} 
27= (3, 3, 1)+(3^{\ast}, 1, 3^{\ast})+(1, 3^{\ast},3)
\ee
where the fields are assigned as follows. $(3, 3, 1)$ corresponds to $(u, d, h)$, $(3^{\ast}, 1, 3^{\ast})$ corresponds to $(h^{c}, d^{c}, u^{c})$ and the leptons are assigned to $(1, 3^{\ast} ,3)$. Here $h$ represents the exotic $-\frac{1}{3}$ charge quark which can carry lepton number depending on the assignments. The other exotic fields are $N^{c}$ and two isodoublets $(\nu_{E}, E)$ and $(E^{c},N_{E}^{c})$. The presence of these exotic fields makes the phenomenology of the low energy subgroups of $E_{6}$ very interesting. The superfields of the first family can be represented as
\be{\label{4.0.1}}
 \bpm u \\ d \\ h \epm + \bpm u^{c} & d^{c} &
h^{c}\epm +\bpm E^{c} & \nu & \nu_{E} \\ N^{c}_{E} & e & E \\
e^{c} & N^{c} & n \epm,
\ee
where $SU(3)_{L}$ operates along columns and $SU(3)_{(R)}$ operates along rows. The $SU(3)_{(L,R)}$ in the maximal subgroup of $E_6$ can further break into $SU(2)_{(L,R)} \times U(1)_{(L,R)}$ and there are three choices of assigning the isospin doublets corresponding to $T, U, V$ isospins (generators of $SU(2)$) of $SU(3)$. One of the choices have $(d^{c}, u^{c})_L$ assigned to the $SU(2)_{R}$ doublet giving rise to the usual left-right symmetric extension of the standard model including the exotic particles. In another choice, the $SU(2)_{R}$ doublet is chosen to be $(h^{c}, d^{c})$ \cite{London:1986dk} with the charge equation given by $Q=T_{3L}+\frac{1}{2} Y_{L}+\frac{1}{2} Y_{N}\,,$ where the chosen $SU(2)_{R}$ does not contribute to the electric charge equation and is often denoted by $SU(2)_{N}$. While these two subgroups are quite interesting from a phenomenological point of view, the superpotential couplings in these two subgroups can not explain the $\cR(D^{(\ast)})$ data. The third possible choice where the $SU(2)_R$ doublet is chosen to be $(h^{c}, u^{c})$ gives the subgroup referred to as the Alternative Left-Right Symmetric Model (ALRSM) \cite{Ma:1986we} and it has the right ingredients to address $\cR(D^{(\ast)})$ excesses. 

In ALRSM, the superfields have the following transformations under the subgroup $G=SU(3)_{c}\times SU(2)_{L}\times SU(2)_{R^{\prime}}\times U(1)_{Y^{\prime}}$  
\bea {\label{4.0.2}}
(u, d)_{L} &:& (3, 2, 1, \frac{1}{6})\nonumber\\
(h^{c}, u^{c})_{L} &:& (\bar{3}, 1, 2, -\frac{1}{6})\nonumber\\
(\nu_{E}, E)_{L} &:& (1, 2, 1, -\frac{1}{2})\nonumber\\
(e^{c}, n)_{L} &:& (1, 1, 2, \frac{1}{2})\nonumber\\
h_{L} &:& (3, 1, 1, -\frac{1}{3})\nonumber\\
d^{c}_{L} &:& (\bar{3}, 1, 1, \frac{1}{3})\nonumber\\
\bpm \nu_{e} & E^{c} \\ e & N^{c}_{E}\epm_{L} &:& (1, 2, 2, 0)\nonumber\\
N^{c}_{L} &:& (1, 1, 1, 0),
\eea
 where $Y^{\prime}=Y_{L}+Y_{R}^{\prime}$. The charge equation is given by $Q=T_{3L}+\frac{1}{2} Y_{L}+T^{\prime}_{3R}+\frac{1}{2} Y^{\prime}_{R}$, where $T^{\prime}_{3R}=\frac{1}{2} T_{3R}+\frac{3}{2}
Y_{R}$, $Y^{\prime}_{R}=\frac{1}{2} T_{3R}-\frac{1}{2} Y_{R}$. The superpotential governing interactions of the superfields in ALRSM is given by \cite{Hewett:1988xc}
 \begin{eqnarray}
 \label{eq:Wcase1}
 && W= \lambda_1\left( u u^{c} N^{c}_E - d u^{c} E^{c} - u h^{c} e + d h^{c} \nu_e \right)+ \nonumber\\
 &&  \lambda_2 \left( u d^{c} E - d d^{c} \nu_{E}\right)+\lambda_3 \left( h u^c e^c - h h^c n\right) + \nonumber\\
 && \lambda_4 h d^c N^{c}_L +\lambda_5 \left ( ee^c \nu_E + E E^c n - E e^{c} \nu_e- \nu_E N^{c}_E n\right) +\nonumber\\
 && \lambda_6 \left( \nu_e N^{c}_L N^{c}_E - e E^c N^{c}_L\right).
 \end{eqnarray}
 The superpotential given in Eq. (\ref{eq:Wcase1}) gives the following assignments of $R$-parity, baryon number ($B$) and lepton number ($L$) for the exotic fermions ensuring proton stability. $h$ is a leptoquark with $R=-1, B=\frac{1}{3}, L=1$. $\nu_{E}, E$ and $n$ have the assignments $R=-1, B=L=0$. $N^{c}$ has two possible assignments. If $N^{c}$ has the assignments $R=-1$ and $B=L=0$ (in a $R$-parity conserving scenario demanding $\lambda_{4}=\lambda_{6}=0$ in Eq. (\ref{eq:Wcase1})), $\nu_{e}$ becomes exactly massless. However if $N^{c}$ is assigned $R=+1$, $B=0$, $L=-1$, then $\nu_{e}$ can acquire a tiny mass via the seesaw mechanism. 
 
 ALRSM can explain both $eejj$ and $e\slashed p_T jj$ signals from the decay of scalar superpartners of the exotic particles, for example, through (i) resonant production of the exotic slepton $\tilde E$, subsequently decaying into a charged lepton and a neutrino, followed by  R-parity conserving interactions of the neutrino producing an excess of events in both  $eejj$ and $e\slashed p_Tjj$ channels \cite{Dhuria:2015hta} (ii) pair production of scalar leptoquarks ${\tilde h}$. On the other hand, high scale leptogenesis can be obtained via the decay of the heavy Majorana neutrino $N^c$ in ALRSM. From the interaction terms $\lambda_{4}$ and $\lambda_{6}$ in Eq. (\ref{eq:Wcase1}), it can be seen that the Majorana neutrino $N^c_{k}$ can decay into final states with $B-L=-1$ given by $\nu_{e_{i}} {\tilde N}^{c}_{E_{j}}, {\tilde \nu}_{e_{i}} N^{c}_{E_{j}}, e_{i}{\tilde E}^{c}_{j}, {\tilde e}_{i}, E^{c}_{j}$ and $d_{i} {\tilde h}_{j}, {\tilde d^{c}}_{i} {\tilde h}_{j}$ and to their conjugate states. Thus, ALRSM has the attractive feature that it can explain both the excess $eejj$ and $e\slashed p_T jj$ signals and also high-scale leptogenesis \cite{Dhuria:2015hta}.
 
 Having introduced ALRSM above now we are ready to analyze the semitauonic $B$ decay $\bar{B}\rightarrow D^{(\ast)} \tau \bar{\nu}$ based on the general framework introduced in Sec. \ref{sec1}. From the superpotential given in Eq. (\ref{eq:Wcase1}) it follows that in ALRSM there are two possible diagrams shown in Fig. \ref{fig1}. which can contribute to the decay $\bar{B}\rightarrow D^{(\ast)} \tau \bar{\nu}$. The effective Lagrangian corresponding to these diagrams is given by

\begin{widetext}
\be
\label{2.1}
\cL_{\rm{eff}}=-\sum_{j,k=1}^{3} V_{2k}\left[ \frac{ \l^{5}_{33j} \l^{2\ast}_{3kj}}{m_{\tilde{E}^j}^{2}} \bar{c}_{L}b_{R} \; \bar{\tau}_{R}\nu_{L} + \frac{ \l^{1}_{33j} \l^{1\ast}_{3kj}}{m_{\tilde{h}^{j \ast}}^{2}} \bar{c}_{L}(\tau^{c})_{R} \; (\bar{\nu}^{c})_{R} b_{L} \right],
\ee
\end{widetext}
where the superscript corresponds to the superpotential coupling index and the generation indices are explicitly written as subscripts. Here $m_{\tilde{E}} (m_{\tilde{h}})$ is the mass of slepton $\tilde{E}^{j}$ (scalar leptoquark $\tilde{h}^{j\ast}$) and $V_{ij}$ corresponds to the $ij$-th component of the CKM matrix. Using Fiertz transformation the second term of Eq. (\ref{2.1}) can be put in the form given by
\be
\label{2.2}
\bar{c}_{L}(\tau^{c})_{R} \; (\bar{\nu}^{c})_{R} b_{L} = \frac{1}{2} \bar{c}_{L}\gamma^{\mu} b_{L} \; \bar{\tau}_{L} \gamma_{\mu} {\nu}_{L}.
\ee
We can now readily obtain the expressions for the corresponding Wilson coefficients, defined in Eq. (\ref{1.1.1}), given by
\bea
\label{2.3}
C^{\tau}_{S_L} &=& \frac{1}{2\sqrt{2} G_{F} V_{cb}} \sum_{j,k=1}^{3} V_{2k} \frac{ \l^{5}_{33j} \l^{2\ast}_{3kj}}{m_{\tilde{E}^j}^{2}} \; ,\nonumber\\
C^{\tau}_{V_L} &=& \frac{1}{2\sqrt{2} G_{F} V_{cb}} \sum_{j,k=1}^{3} V_{2k} \frac{ \l^{1}_{33j} \l^{1\ast}_{3kj}}{2\,m_{\tilde{h}^{j \ast}}^{2}},
\eea
where the neutrinos are assumed to be predominantly of tau flavor.

To simplify further analysis, we invoke the assumption that except the SM contribution only one of the NP operators in Eq. (\ref{1.1.1}) contributes dominantly. This assumption helps us in determining the limits on the dominant Wilson coefficient from the experimental data for $\cR(D^{(\ast)})$ and the generalization of this situation to incorporate more than one NP operator contribution is straight forward. 

\begin{figure}[ht!]
   \hspace{0.02cm}
    \hbox{\hspace{0.03cm}
    \hbox{\includegraphics[scale=0.25]{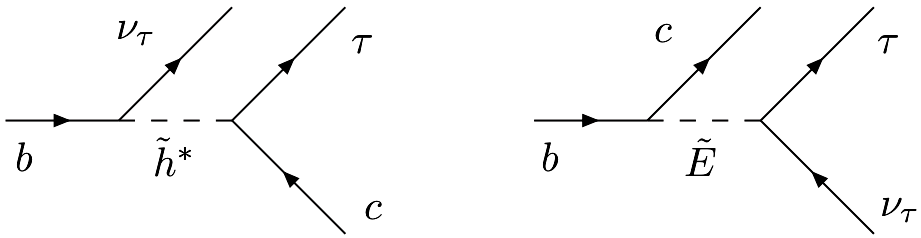}}
    }
    
    \caption{ Feynman diagrams for the decays $\bar{B}\rightarrow D^{(\ast)} \tau \bar{\nu}$ induced by the exchange of scalar leptoquark ($\tilde{h}^{\ast}$) and $\tilde{E}$.
     }
     \label{fig1}
    \end{figure}

The case where $C^{\tau}_{S_L}$ is the dominant contribution, similar to 2HDM of type II or type III with minimal flavor violation, can not explain both $\cR(D)$ and $\cR(D^{\ast})$ data simultaneously \cite{Fajfer:2012jt, Crivellin:2012ye}, as can be seen from Fig. \ref{fig2a}. However, $C^{\tau}_{V_L}$ has an allowed region which can explain both $\cR(D)$ and $\cR(D^{\ast})$ data as shown in Fig. \ref{fig2b}. We find that for $\left| C^{\tau}_{V_{L}} \right| > 0.08$ the current experimental data can be explained. Note that, we use the central values of the theoretical predictions because the theoretical uncertainties are sufficiently small compared to the experimental accuracy.
 \begin{figure}[ht!]
   \hspace{0.02cm}
    \hbox{\hspace{0.03cm}
    \hbox{\includegraphics[scale=0.35]{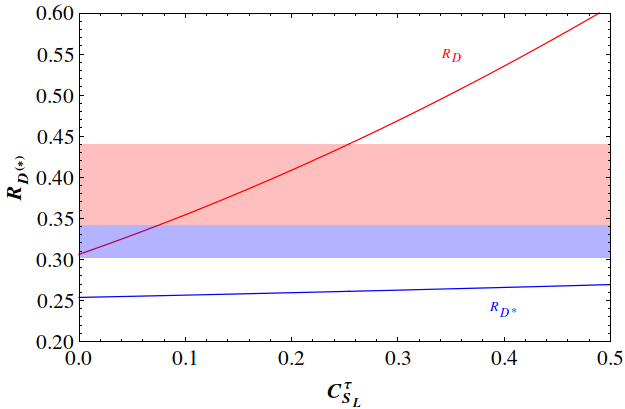}}
    }
    
    \caption{ The dependence of the observables $R_{D^{(*)}}$ on $C_{S_L}^\tau$: red (blue) line corresponds to $R_D$ ($R_{D^{*}}$), and the horizontal light red (blue) band corresponds to the experimentally allowed $1\sigma$ values. No common region exists for $C_{S_L}^\tau$ which can simultaneously explain both $R_D$ and $R_{D^{*}}$.
     }
     \label{fig2a}
    \end{figure}
    
     \begin{figure}[ht!]
   \hspace{0.02cm}
    \hbox{\hspace{0.03cm}
    \hbox{\includegraphics[scale=0.35]{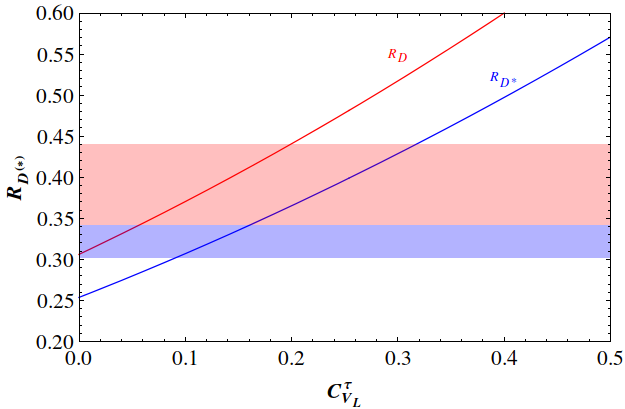}}
    }
    
    \caption{ The dependence of the observables $R_{D^{(*)}}$ on $C_{V_L}^\tau$: red (blue) line corresponds to $R_D$ ($R_{D^{*}}$), and the horizontal light red (blue) band corresponds to the experimentally allowed $1\sigma$ values. $C_{V_L}^\tau$ can explain both $R_D$ and $R_{D^{*}}$ data. 
     }
     \label{fig2b}
    \end{figure}
\section{Constraints from $B$, $D$ decays and $D^{0}-\bar{D}^{0}$ oscillation}
\label{sec3}
\subsection{Constraints from $B \rightarrow \tau \nu $}
In this section we discuss the new contributions to purely leptonic decay mode $B \rightarrow \tau \nu $ due to scalar leptoquark $\tilde{h}^{j\ast}$ exchange and utilize the measured branching fractions of the decay to derive constraints on the product of  couplings $\l^{1}_{33j} \l^{1}_{31j}$. In the SM, the decay $B \rightarrow \tau \nu $ proceeds via annihilation to a W boson in the s-channel. In the ALRSM, the exchange of the scalar leptoquark $\tilde{h}^{j\ast}$ leads to the additional diagrams shown in Fig. \ref{fig3}. Since the mass scale of scalar leptoquark is far above the scale of the B meson, we can integrate out the heavy degree of freedom to generate new four-fermion interaction $\sim\bar{q}_{L}(\tau^{c})_{R} \; (\bar{\nu}^{c})_{R} b_{L}$, with the Wilson coefficients parameterizing the effects of the integrated out non-standard particles. The NP effective Hamiltonian is given by
\bea
\label{4.1.1}
\cH_{\rm{eff}}^{\rm{NP}}(b\bar{q}\rightarrow\tau\bar{\nu})= \frac{4 G_F}{\sqrt{2}} V_{qb} \; C^{qb}_{V_L} (\bar{q}_L\gamma^\mu b_L)(\bar{\tau}_L\gamma_\mu \nu_L),
\eea
where $V_{qb}$ (here $q\equiv u$) is the relevant CKM matrix element. The Wilson coefficient $C^{ub}_{V_L}$ in terms of the couplings $\l'$s is given by
\bea
\label{4.1.2}
C^{ub}_{V_L} &=& \frac{1}{2\sqrt{2} G_{F} V_{ub}} \sum_{j,k=1}^{3} V_{1k} \frac{ \l^{1}_{33j} \l^{1\ast}_{3kj}}{2\,m_{\tilde{h}^{j \ast}}^{2}}.
\eea
 \begin{figure}[ht!]
   \hspace{0.02cm}
    \hbox{\hspace{0.03cm}
    \hbox{\includegraphics[scale=0.3]{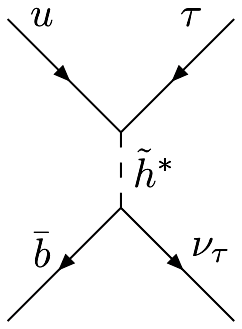}}
    }
    
    \caption{ Feynman diagrams for the decay $B \rightarrow \tau \nu $ induced by the exchange of the scalar leptoquark $\tilde{h}^{j\ast}$.
     }
     \label{fig3}
    \end{figure}
In our notation, the Wilson coefficient of the SM effective operator is set to unity. In what follows, we will neglect the subleading $\cO(\l)$ terms and retain only the leading CKM element $V_{11}$. \\
Note that, the decay $B \rightarrow \tau \nu$ is the only experimentally measured purely leptonic mode of charged $B^{\pm}$. The current experimental value of the branching ratio of  $B \rightarrow \tau \nu$ is $(1.14 \pm 0.27) \times 10^{-4}$ \cite{Agashe:2014kda}. The presence of NP modifies the expression of the SM decay rate in the following way

\bea
\label{4.1.3}
\frac{d\Gamma}{dq^2}(B\rightarrow \tau \nu) =&& \frac{G_F^2 |V_{ub}|^2}{8 \pi}m_B f_B^2 m_\tau^2 \nonumber \\ 
&& \times \left(1-\frac{m_\tau^2}{m_B^2}\right)^2 \lvert1 + C_{V_L}^{ub}\rvert^2 ,
\eea
where $m_{B}$ is the mass of $B^\pm$ and $f_{B}$ is the decay constant which parametrize the matrix elements of the corresponding current as
\bea
\label{4.1.4}
\langle 0 \lvert \bar{b}_L \gamma^\mu  q_L \rvert B_q(p_B)\rangle = p_B^\mu f_{B}.
\eea
Here $p_{B}$ is the 4-momentum of the $B^\pm$ meson.\\
We use the CKM matrix elements, the lifetimes, particle masses and decay constants $f_B$, $f_{D_s}$, $f_{D^+}$ from PDG \cite{Agashe:2014kda} for numerical estimations throughout the paper. There have been attempts to account for flavour symmetry breaking in pseudoscalar meson decay constants in literature \cite{Gershtein:1976mv, Khlopov:1978id}.  Here, we assume that contribution from only one type of scalar leptoquarks is dominant and real. For simplicity, we will further assume the couplings to be real in the rest of this paper. In Fig. \ref{fig4} we plot the BR($B \rightarrow \tau \nu$) as a function of the product of the couplings $\l_{33j} \l_{31j}$ for different values of $m_{\tilde{h}^{j \ast}}$. Numerically these constraints are given by
\bea
\label{4.1.5}
\l_{33j} \l_{31j}\leq 0.04 \left( \frac{m_{\tilde{h}^{j \ast}}}{1000 \rm{GeV}}\right)^2.
\eea

    \begin{figure}[ht!]
  \hspace{0.02cm}
   \hbox{\hspace{0.03cm}
   \hbox{\includegraphics[scale=0.37]{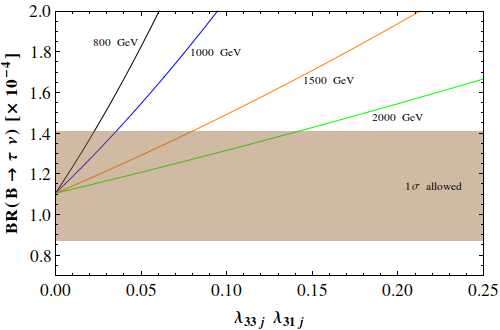}}
   }
   \caption{ BR($B \rightarrow \tau \nu$) as a function of couplings $\l_{33j} \l_{31j}$ for $m_{\tilde{h}^{j \ast}} = $ 800, 1000, 1500, 2000 GeV corresponding to black, blue, orange, and green lines respectively. The horizontal brown (light) band shows the $1\sigma$ experimentally favored values.
    }
    \label{fig4}
   \end{figure}

\subsection{Constraints from $D_s^+ \rightarrow \tau \nu $ and $D^+ \rightarrow \tau \nu $}
Along with rare B decays, the study of the decays of charmed mesons also offer attractive possibilities to test the predictions of extensions of the SM \cite{Buchalla:2008jp, Burdman:2001tf}. In fact, these processes are quite sensitive to the contributions of charged Higgs boson and scalar leptoquarks \cite{Dobrescu:2008er} and to the new contributions from squark exchange in the framework of R-parity violating SUSY as examined in Ref. \cite{Kao:2009fg}. In this section we consider the purely leptonic decays $D_s^+ \rightarrow \tau \nu $ and $D^+ \rightarrow \tau \nu $ in ALRSM and use their measured branching ratios to obtain  constraints on the couplings $(\l_{32j})^2$ and $\l_{32j} \l_{31j}$ respectively. The relevant Feynman diagrams in ALRSM for the decays  $D_s^+ \rightarrow \tau \nu $ and $D^+ \rightarrow \tau \nu $ are shown in Fig. \ref{fig5}. Integrating out the heavy energy scales yields the following non-standard effective Hamiltonian
\bea
\label{4.2.1}
\cH_{\rm{eff}}^{\rm{NP}}(c\bar{q}\rightarrow\tau\bar{\nu})= \frac{4 G_F}{\sqrt{2}} V_{cq} \; C^{cq}_{V_L} (\bar{q}_L\gamma^\mu c_L)(\bar{\nu}_L\gamma_\mu \tau_L)
\eea
where $q = s, d$ for $D_s^+, D^+$ respectively. In the SM these processes occur (similar to $B\rightarrow \tau \nu$) via 
$W^\pm$ annihilation in the s-channel  and the SM Wilson coefficient is given by unity in our notation. The corresponding Wilson coefficient $C^{cq}_{V_L}$ parameterizing the NP effects is given by
\bea
\label{4.2.2}
C^{cq}_{V_L} &=& \frac{1}{2\sqrt{2} G_{F} V_{cq}} \sum_{j,k=1}^{3} V_{kq} \frac{ \l^{1}_{32j} \l^{1\ast}_{3kj}}{2\,m_{\tilde{h}^{j \ast}}^{2}}.
\eea
 \begin{figure}[ht!]
   \hspace{0.02cm}
    \hbox{\hspace{0.03cm}
    \hbox{\includegraphics[scale=0.3]{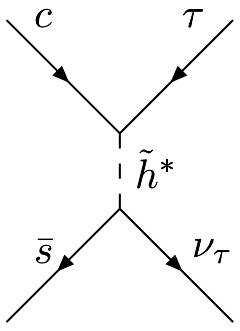}}
    }
    
    \caption{ Feynman diagrams for the decay  $D_s^+ \rightarrow \tau \nu $ induced by scalar leptoquarks. The corresponding diagram for the decay $D^+ \rightarrow \tau \nu $ can be obtained by replacing $s$ quark by $d$ quark.
     }
     \label{fig5}
    \end{figure}
We will keep only the leading terms $V_{cs}$ for $D_s^+$ decay and $V_{ud}$ for $D^+$ case respectively and neglect the subleadiing Cabibbo suppressed $\cO(\l)$ terms. Although this process occurs in the SM at the tree level, the branching fraction is helicity-suppressed. For $\tau$, this suppression is less severe but phase-space suppression is larger compared to light leptons. In the presence of scalar leptoquark contribution, the SM decay rate is  affected in the following way \cite{Dobrescu:2008er, Barranco:2013tba}
 \bea
 \label{4.2.3}
 \frac{d\Gamma}{dq^2}(D_q^+\rightarrow \tau \nu) =&& \frac{G_F^2 |V_{cq}|^2}{8 \pi}m_{D_q} f_{D_q}^2 m_\tau^2  \nonumber\\ 
 && \times \left(1-\frac{m_\tau^2}{m_{D_q}^2}\right)^2 \lvert1 + C_{V_L}^{cq}\rvert^2 .
 \eea
 Here $m_{D_q}$ is the mass of charm-mesons $D_s^+$ and $D^+$ for $q=s,d$ respectively and $V_{cq}$ is the relevant CKM element. The decay constant $f_{D_q}$ is defined by $\langle 0 \lvert \bar{c}_L \gamma^\mu  q_L \rvert D_q(p_{D_q})\rangle = p_{D_q}^\mu f_{D_q}$, where $p_{D_q}$ is the 4-momentum of the $D_q$ meson.\\
Assuming that only one product combination of the scalar leptoquark couplings is nonzero, we get upper bounds on $(\l^{1}_{32j})^2$ and $\l^{1}_{32j} \l^{1\ast}_{31j}$. In Fig. \ref{fig6} we plot the dependence of BR($B \rightarrow {D_(s)}^{+} \nu$) on the coupling $\l_{32j} \l_{31j} (\l_{32j}^2)$ for different $m_{\tilde{h}^{j \ast}}$. Numerically the constrains are given by
\bea
\label{4.2.4}
\l_{32j}^2 \leq 0.85 \left( \frac{m_{\tilde{h}^{j \ast}}}{1000 \rm{GeV}}\right)^2, \nonumber\\
\l_{32j} \l_{31j}\leq 3.12 \left( \frac{m_{\tilde{h}^{j \ast}}}{1000\rm{GeV}}\right)^2.
\eea
As discussed in the next subsection, we find that a more constraining bound on the product of the couplings  $\l_{32j} \l_{31j}$ can be obtained from $D^{0}-\bar{D}^{0}$ mixing as compared to those obtained from $D^+ \rightarrow \tau \nu $.
     \begin{figure}[ht!]
   \hspace{0.02cm}
    \hbox{\hspace{0.03cm}
    \hbox{\includegraphics[scale=0.37]{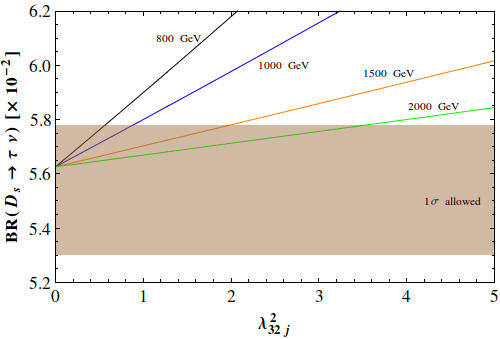}}
    }
     \hbox{\hspace{0.03cm}
        \hbox{\includegraphics[scale=0.37]{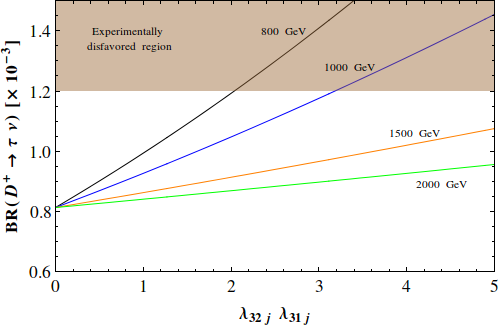}}
        }
    \caption{ Dependence of (upper figure) BR($D_s^{+} \rightarrow \tau \nu$) on the coupling $\l_{32j}^2$ [(lower figure)BR($D^{+} \rightarrow \tau \nu$) on the coupling $\l_{32j} \l_{31j}$] for $m_{\tilde{h}^{j \ast}} = $ 800, 1000, 1500, 2000 GeV corresponding to black, blue, orange, and green lines respectively. In the upper (lower) figure the horizontal brown band shows the $1\sigma$ experimentally allowed (disfavored) region.
     }
     \label{fig6}
    \end{figure}
\subsection{Constraints from $D^0-\bar{D}^0$ mixing}
The phenomenon of meson-antimeson oscillation, being a flavor changing neutral current (FCNC) process, is very sensitive to heavy particles propagating in the mixing amplitude and therefore, it provides a powerful tool to test the SM and a window to observe NP.  In the $D^0-\bar{D}^0$ system, $b$-quark contribution to the fermion loop of the box diagram provides a $\Delta C=2$ transition which is highly suppressed $\sim\cO{(\l^3)}$ (by a tiny $V_{ub}$ CKM matrix element). Therefore, the large non-decoupling effects from a heavy fermion in the leading one-loop contributions is small. $D^0-\bar{D}^0$ mixing involves the dynamical effects of rather light down-type particles and therefore it provides information complementary to the strange and bottom systems where the large effects of heavy top quark in the loops are quintessential. The $D^0-\bar{D}^0$ mixing is described by $\D C = 2$ effective Hamiltonian which induces off-diagonal terms in the mass matrix for neutral D meson pair  and typically parametrized in terms of following experimental observables
\bea\label{4.2.4}
x_D \equiv \frac{\D M_D}{\G_D} \;  \rm{and}\; \;\; y_D \equiv \frac{\G_D}{2\G_D},\nonumber\\
\eea
where $\D M_D$ and $\D \G_D$ are the mass and width splittings between mass eigenstates of  $D^0-\bar{D}^0$ systems respectively and $\G_D$ is the average width. The parameters $x_D$ and $y_D$ can be written in terms of the mixing matrix as follows
\bea
\label{4.2.5}
x_D &=& \frac{1}{2 M_D \G_D} \rm{Re}\left[2\langle\bar{D}^0\lvert H^{|\D C|=2}\rvert D^0\rangle \right. \nonumber\\
&& \left.  + \langle\bar{D}^0\lvert i \int d^4x T\{\cH_w^{|\D C|=1}(x) \cH_w^{|\D C|=1}(0)\}\rvert D^0\rangle  \right],\nonumber\\
y_D &=& \frac{1}{2 M_D \G_D} \rm{Im}\langle\bar{D}^0\lvert i \int d^4x\nonumber\\
 &&\; \times \;\; T\{\cH_w^{|\D C|=1}(x) \cH_w^{|\D C|=1}(0)\}\rvert D^0\rangle,
\eea 
with $\cH_w^{|\D C|=1}(x)$ being the Hamiltonian density that describes $|\D C| =1$ transitions at space-point $x$ and T denotes the time ordered product. 
Since the local $|\D C| =2$ interaction does not contain an absorptive part, this term does not affect $y_D$ and contributes to $x_D$ only. The measured values of $x_D$ and $y_D$ as determined by HFAG are \cite{HFAG:mixing parameters}
\bea\label{4.2.6}
x_D &=& 0.49^{+0.14}_{-0.15} \times 10^{-2},\nonumber\\
y_D &=& (0.61 \pm 0.08) \times 10^{-2},
\eea
 \begin{figure}[ht!]
   \hspace{0.02cm}
    \hbox{\hspace{0.02cm}
    \hbox{\includegraphics[scale=0.2]{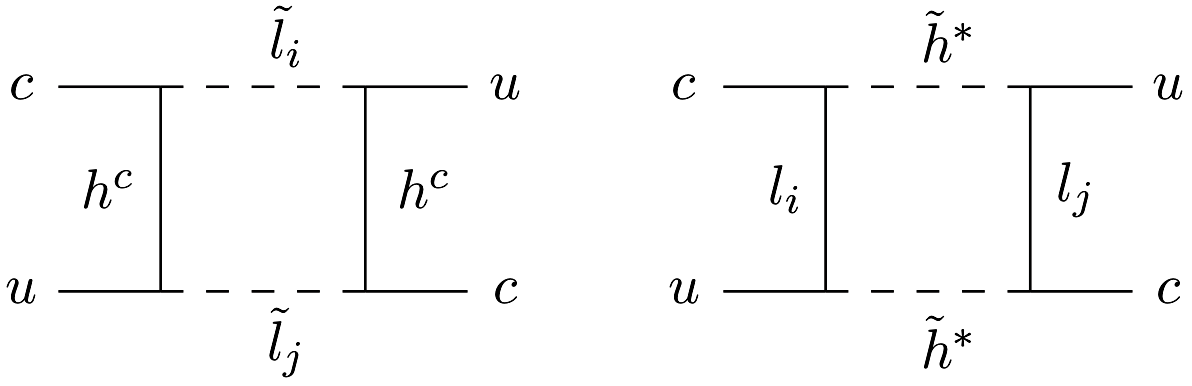}}
    }
    
    \caption{ Feynman diagrams contributing to $D^{0}-\bar{D}^{0}$ mixing in ALRSM induced by scalar leptoquark and slepton.
     }
     \label{fig7}
    \end{figure}
Charm mixing in the SM is highly affected by contributions from intermediate hadronic states, and therefore the theoretical estimations in the SM suffers from large uncertainties and  generally stretched over several orders of magnitude (for a review, see Ref. \cite{Golowich:2007ka}). Like in the case of mixing in neutral K and B systems,   $D^0-\bar{D}^0$ mixing is also sensitive to NP effects. Both $x_D$ and $y_D$ can receive large contributions from NP. The contribution to $y_D$ in several NP models including LR models, multi Higgs models, SUSY without R-parity violations and models with extra vector like quarks has been studied in Ref. \cite{Golowich:2006gq}, while in Ref. \cite{Golowich:2007ka} the NP contributions to $x_D$ in 21 NP models have been discussed. In this section, we use the neutral D meson mixing to obtain constraints on $\l_{32j} \l_{31j}$. These bounds are more tighter than those obtained in the previous section from measured BR of  $D^+ \rightarrow \tau \nu $. The relevant Feynman diagrams which contribute to $D^0-\bar{D}^0$ mixing in the ALRSM are shown in Fig. \ref{fig7}. These Box diagrams are similar to the diagrams generated from internal line exchange of lepton-squark pair or slepton-quark pair in the case of R-parity violating models \cite{Agashe:1995qm, Golowich:2007ka}. The mixing is described by the effective Hamiltonian
\bea\label{4.2.7} 
\cH_{\rm{eff}} &=& \frac{1}{128 \pi^2}(\l_{32j}\l_{31j})^2\left(\frac{1}{m^2_{\tilde{\tau}}} + \frac{1}{m^2_{\tilde{h}^{j \ast}}} \right)\nonumber\\
&& \;\;\;\;\times \;\;(\bar{c}_L\gamma^\mu u_L)(\bar{c}_L\gamma_\mu u_L),
\eea
where we assume that the box diagrams receive contributions from third generation of leptons only. Following Ref. \cite{Agashe:1995qm, Golowich:2007ka} and taking $m_{\tilde{h}^{j \ast}} \simeq m_{\tilde{\tau}}$, the constraints on the size of couplings is given by
\bea\label{4.2.8}
\l_{32j}\l_{31j} \leq 0.17 \sqrt{x_D^{\rm{expt}}} \left(\frac{m_{\tilde{h}^{j \ast}}}{1000 \rm{GeV}}\right).
\eea
In Fig. \ref{fig8}, we plot the dependence of $x_D^{ALRSM}$ on the product of the couplings $\l_{32j}\l_{31j}$ for different $m_{\tilde{h}^{j \ast}}$.
     \begin{figure}[ht!]
   \hspace{0.02cm}
    \hbox{\hspace{0.03cm}
    \hbox{\includegraphics[scale=0.49]{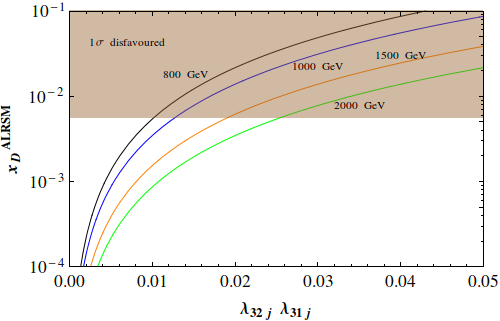}}
    }
    
    \caption{ Dependence of $x_D^{ALRSM}$ on the coupling $\l_{32j}\l_{31j}$ for $m_{\tilde{h}^{j \ast}} = $ 800, 1000, 1500, 2000 GeV corresponding to black, blue, orange, and green lines respectively. The horizontal brown (light) band shows the $1\sigma$ experimentally disfavored region.
     }
     \label{fig8}
    \end{figure}

\section{Results and discussion}
\label{sec4}
Having discussed the allowed region for $C^{\tau}_{V_L}$ which can explain both $\cR(D)$ and $\cR(D^{\ast})$ data simultaneously in Sec. \ref{sec2} and the constraints on the couplings $\l_{33j}$ and $\l_{32j}$  involved in $C^{\tau}_{V_L}$ from the leptonic decays $D_{s}^{+} \rightarrow \tau^{+} \bar{\nu}$, $B^{+}\rightarrow \tau^{+} \bar{\nu}$, $D^{+}\rightarrow \tau^{+} \bar{\nu}$ and $D^{0}$-$\bar{D}^{0}$ mixing in Sec. \ref{sec3}, we are now ready to translate these analysis into a simple $\l_{33j}$-$\l_{32j}$ parameter space analysis.
    \begin{figure}[ht!]
   \hspace{0.02cm}
    \hbox{\hspace{0.03cm}
    \hbox{\includegraphics[scale=0.35]{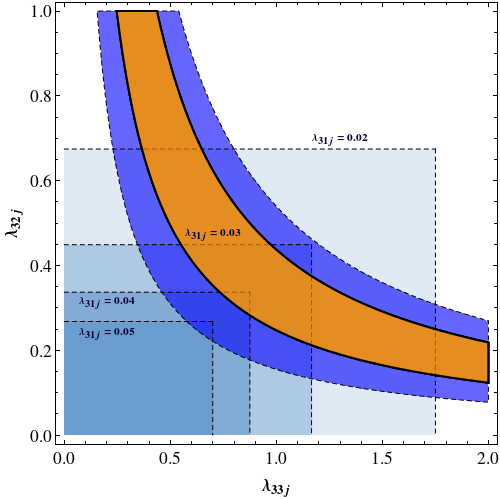}} }
     \caption{ 
     The region of $\l_{33j}$-$\l_{32j}$ parameter space compatible with the experimental data for $\cR(D^{(\ast)})$ and constraints from the leptonic decays $D_{s}^{+} \rightarrow \tau^{+} \bar{\nu}$, $B^{+}\rightarrow \tau^{+} \bar{\nu}$, $D^{+}\rightarrow \tau^{+} \bar{\nu}$ and $D^{0}$-$\bar{D}^{0}$ mixing. We take $m_{\tilde{h}^j\ast}=1000 \gev$ for this plot. Blue band between dashed lines shows allowed values considering constraints from $R_D$ only, Orange band between bold black lines shows allowed region favored by experimental data for both $R_{D^{*}}$ and $R_D$. The shaded (light blue) rectangles correspond to the allowed regions of $\l_{33j}$-$\l_{32j}$ parameter space for different values of $\l_{31j}$ marked with the corresponding allowed upper boundary shown in dashed lines consistent with the present experimental data on $B\rightarrow \tau \nu$, $D_s\rightarrow \tau \nu$, $D^{+}\rightarrow \tau \nu$ and $D-\bar{D}$ mixing.  
     }
     \label{fig_fin}
    \end{figure}
In Fig. \ref{fig_fin}, we plot the range of the couplings $\l_{33j}$ and $\l_{32j}$ (for $m_{\tilde{h}^j\ast}=1000 \gev$) that can explain both $\cR(D)$ and $\cR(D^{\ast})$ data over the parameter space allowed by the the leptonic decays and $D^{0}$-$\bar{D}^{0}$ mixing. From the decay $D_{s}^{+} \rightarrow \tau^{+} \bar{\nu}$, we constrain the allowed upper limit of the coupling $\lambda_{32j}$. The decay  $D^{+}\rightarrow \tau^{+} \bar{\nu}$ and $D^{0}$-$\bar{D}^{0}$ mixing give constraints on the upper limit of the product of couplings $\l_{32j}\l_{31j}$. We find that among the two processes the latter gives more stringent constraints and therefore we use the constrains on the allowed upper limit of $\l_{32j}\l_{31j}$ coming from $D^{0}$-$\bar{D}^{0}$ mixing. Finally, we use the decay $B^{+}\rightarrow \tau^{+} \bar{\nu}$ to constrain the upper limit of $\l_{33j}\l_{31j}$. The latter two constraints on the products of couplings have $\l_{31j}$ as a common free parameter and the shaded rectangles in Fig. \ref{fig_fin} correspond to the allowed regions of $\l_{33j}$-$\l_{32j}$ parameter space for different values of $\l_{31j}$ marked in the figure with the corresponding allowed upper boundary shown in dashed lines. The blue band corresponds to the allowed band of $\l_{33j}$-$\l_{32j}$ explaining the $\cR(D)$ data and the orange band corresponds to the allowed band of $\l_{33j}$-$\l_{32j}$ explaining both $\cR(D)$ and $\cR(D^{\ast})$ data simultaneously. We would like to note that the list of constraints mentioned above is far from exhaustive and many other possible leptonic and semileptonic decays can give independent constrains. For instance, the decay process $\tau^{+} \rightarrow \pi^{+} \nu$ can give independent constraint on $\lambda_{31j}$, which we find to be consistent with the values extracted out of the above constraints and used for the parameter space analysis. On the other hand, the semileptonic decay $t\rightarrow b \tau \nu$ can give constraint on $\l_{33j}$ which we find to be again consistent with the values used in the above parameter space analysis.  Also the effective NP operators under consideration may induce $B$-decays such as $b\rightarrow s\nu \bar{\nu}$ \cite{Grossman:1995gt, Buras:2014fpa}, which can be an interesting channel for the future experiments.

In conclusion, we have studied the superstring inspired $E_6$ motivated Alternative Left-Right Symmetric model to explore if this model can explain the current experimental data for both $\cR(D)$ and $\cR(D^{(\ast)})$ simultaneously addressing the excesses over the SM expectations. We use the leptonic decays $D_{s}^{+} \rightarrow \tau^{+} \bar{\nu}$, $B^{+}\rightarrow \tau^{+} \bar{\nu}$, $D^{+}\rightarrow \tau^{+} \bar{\nu}$ and $D^{0}$-$\bar{D}^{0}$ mixing to constrain the couplings involved in the semileptonic $b \rightarrow c$ transition in ALRSM. We find that ALRSM can explain the current experimental data on $\cR(D^{(\ast)})$ quite well while satisfying the constraints from the rare $B$, $D$ decays $D^{0}$-$\bar{D}^{0}$ mixing. Furthermore, ALRSM can also explain both the $eejj$ and $e {\slashed p_T} jj$ signals recently reported by CMS and also accommodate successful leptogenesis. If these excess signals are confirmed in future B-physics experiments and at the LHC then ALRSM will be an interesting candidate for NP beyond the Standard Model.

\section*{Acknowledgment}
CH would like to thank Utpal Sarkar for many helpful discussions.  
 
\end{document}